\documentclass[prb,aps,superscriptaddress,reprint]{revtex4-1}

\usepackage{graphicx}
\usepackage[english]{babel}
\usepackage{amsmath}
\usepackage{SIunits}
\usepackage[utf8]{inputenc}
\usepackage{xcolor}

\newcommand{\be}{\begin{equation}}
\newcommand{\ee}{\end{equation}}
\newcommand{\bea}{\begin{eqnarray}}
\newcommand{\eea}{\end{eqnarray}}

\newcommand{\Yambo} {{\normalfont\ttfamily Yambo}}

\def\gL         {\Lambda}

\def\bverb      {\renewcommand{\baselinestretch}{1}\begin{verbatim}}
\def\everb  	{\end{verbatim}\renewcommand{\baselinestretch}{1.3}}

\def\gL         {\Lambda}

\newcommand{\etsf} {European Theoretical Spectroscopy Facilities (ETSF)}

\newcommand{\liege}{nanomat/Q-mat/CESAM, Universit\'e de Li\`ege, B-4000 Sart Tilman, Li\`ege, Belgium}
\newcommand{\barcelona}{Catalan Institute of Nanoscience and Nanotechnology (ICN2), CSIC and BIST, Campus UAB, Bellaterra, 08193 Barcelona, Spain}

\newcommand{\aachenphys}{2nd Institute of Physics and JARA-FIT, RWTH Aachen University, D-52074 Aachen, Germany}

\newcommand{\pgi}{Peter Gr\"unberg Institute (PGI-9), Forschungszentrum J\"ulich, D-52425 J\"ulich, Germany}

\begin{document} 

\title{Spin States Protected from Intrinsic Electron-Phonon-Coupling Reaching 100 ns Lifetime at Room Temperature in MoSe$_2$}

\author{Manfred Ersfeld}
\affiliation{\aachenphys}
\author{Frank Volmer}
\affiliation{\aachenphys}
\author{Pedro Miguel M. C. de Melo}
\affiliation{\liege}
\affiliation{\etsf}
\author{Robin de Winter}
\affiliation{\aachenphys}
\author{Maximilian Heithoff}
\affiliation{\aachenphys}
\author{Zeila Zanolli}
\affiliation{\etsf}
\affiliation{\barcelona}
\affiliation{Institute for Theoretical Solid State Physics,
RWTH Aachen University, D-52056 Aachen, Germany}
\author{Christoph Stampfer}
\affiliation{\aachenphys}
\affiliation{\pgi}
\author{Matthieu J. Verstraete}
\affiliation{\liege}
\affiliation{\etsf}
\author{Bernd Beschoten}
\affiliation{\aachenphys}
\email{bernd.beschoten@physik.rwth-aachen.de}




\begin{abstract}
We present time-resolved Kerr rotation measurements, showing spin lifetimes of over \unit{100}{ns} at room temperature in monolayer MoSe$_2$. These long lifetimes are accompanied by an intriguing temperature dependence of the Kerr amplitude, which increases with temperature up to \unit{50}{K} and then abruptly switches sign. Using \textit{ab initio} simulations we explain  the latter behavior in terms of the intrinsic electron-phonon coupling and the activation of transitions to secondary valleys. The phonon-assisted scattering of the photo-excited electron-hole pairs prepares a valley spin polarization within the first few ps after laser excitation. The sign of the total valley magnetization, and thus the Kerr amplitude, switches as a function of temperature, as conduction and valence band states exhibit different phonon-mediated inter-valley scattering rates. However, the electron-phonon scattering on the ps time scale does not provide an explanation for the long spin lifetimes. Hence, we deduce that the initial spin polarization must be transferred into spin states which are protected from the intrinsic electron-phonon coupling, and are most likely resident charge carriers which are not part of the itinerant valence or conduction band states.
\end{abstract}

\maketitle

Monolayers of transition metal dichalcogenides (TMDs) like molybdenum diselenide (MoSe$_2$) are two-dimensional (2D) semiconductors with a direct band gap and a strong spin orbit splitting of the valence band of several \unit{100}{meV}.\cite{PhysRevB.84.153402} Optical selection rules allow a valley selective excitation of spin states with circular polarized light, which makes this class of material interesting for spintronic applications.\cite{PhysRevLett.108.196802, NatPhys.10.343,PhysRevB.92.155403} The performance of TMDs for spintronics is backed by the measurement of extraordinarily long spin lifetimes in the $\mu$s range both in TMD monolayers and heterostructures at cryogenic temperatures.\cite{PhysRevLett.119.137401, Science.360.893, Kim2017Jul} Nevertheless, realizing the technological potential of monolayer TMDs will require technologically relevant spin properties at room temperature (RT). In this respect, already the archetype 2D-material, graphene, has demonstrated that it is possible to outperform conventional spintronic materials such as GaAs and Si.\cite{2DMaterials.Roche} In the case of graphene, spin lifetimes up to \unit{12.6}{ns} were measured in all-electrical spin precession measurements at RT.\cite{NanoLett.16.3533} In TMDs, optical Kerr rotation measurements revealed that a spin signal with a lifetime of hundreds of ps can survive up to RT.\cite{NatureComm.6.896}

Here, we present time-resolved Kerr rotation (TRKR) measurements on exfoliated monolayer MoSe$_2$ flakes, showing room temperature spin lifetimes up to \unit{100}{ns}. By the combination of energy-dependent, two-color pump-probe Kerr measurements, and photoluminescence spectroscopy on samples with different doping and defect levels,
we show that these long lifetimes are likely linked to resident carriers which are not itinerant states. \textit{ab initio} calculations of electron-phonon coupling show that itinerant carriers are expected to undergo fast phonon-induced spin relaxation. 
We show that phonon-induced scattering rates of valence and conduction band states are the key to understand the formation of an initial valley magnetization, which is directly probed by the Kerr rotation. An intriguing increase of the Kerr rotation amplitude, followed by a sign reversal at higher temperatures, can be explained by a temperature-activated change of the conduction and valance band scattering rates with phonons. We argue that the initial laser-prepared magnetization is transferred within ps to long-lived spin states, which are then probed on the ns time scale by Kerr rotation measurements.

\begin{figure*}[tb]
		\includegraphics[width=\linewidth]{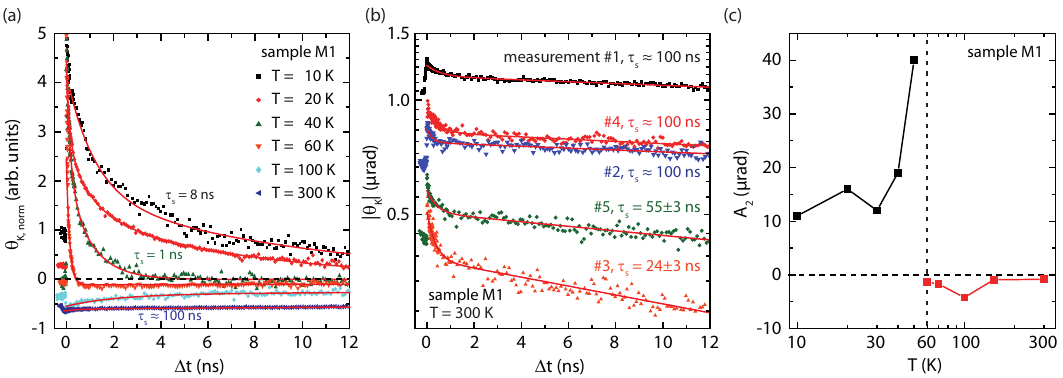}
	\caption{(a) TRKR data of one MoSe$_2$ monolayer (sample M1) at different temperatures, with bi-exponential fits (red curves), showing spin lifetimes of \unit{100}{ns} at room temperature (purple curve). While these long-lived spin states appear above \unit{60}{K} with a Kerr angle of $\Theta_\text{K}<0$, there are different spin states with $\Theta_\text{K}>0$ at lower temperatures with strongly temperature dependent spin lifetimes. We note that all traces have normalized amplitudes to visualize the overall change in lifetime more clearly. (b) TRKR curves measured at RT on different days yield varying lifetimes from \unit{24}{ns} to \unit{100}{ns}. Due to the logarithmic representation of the data, we plot $|\Theta_\text{K}|$. (c) The amplitude $A_2$ of the long-lived spin signal from the bi-exponential fit shows an intriguing initial increase with temperature, before it abruptly switches sign at \unit{60}{K}.}
	\label{figure1}
\end{figure*}

The MoSe$_2$ flakes were mechanically exfoliated from bulk crystals from different suppliers with a polydimethylsiloxane (PDMS) membrane and transferred onto Si/SiO$_2$ substrates.\cite{2DMaterials.1.011022,2DMaterials.4.025030} For the TRKR experiments we use two mode-locked Ti:sapphire lasers to independently tune the energies of both pump and probe pulses. An electronic delay between both pulses covers the full laser repetition interval of \unit{12.5}{ns}. The pulse widths are on the order of \unit{3}{ps}, the laser spot sizes have a FWHM value of approximately $\unit{6-8}{\mu m}$ and the laser power was kept to $\unit{600}{\mu W}$ for both pump and probe beams. The probe energy was set near the trion peak position, determined for each sample and temperature by photoluminescence measurements, whereas the pump energy was set to a slightly higher energy if not noted otherwise. A detailed scheme of the experimental setup can be found in Ref.~\citenum{PhysRevB.95.235408}. 

Fig.~1(a) depicts the TRKR traces for the sample showing the unprecedented long spin lifetimes at higher temperatures (sample M1). A spin polarization is created by the circularly polarized pump pulse while the linearly polarized probe pulse measures its temporal decay. The corresponding Kerr rotation $\Theta_{K}(t)$ is fitted by a bi-exponential fit function of the form: 

\begin{equation}
\Theta_{K}(t) =\sum_{i=1}^2 A_i \cdot \exp \left( -\frac{t}{\tau_\text{s,i}} \right),
\label{eq:Fit}
\end{equation}
 where $A_{i}$ are the amplitudes and $\tau_\text{s,i}$ are the respective spin lifetimes. This accounts for a rapid initial depolarization which occurs on a time scale of tens to hundreds of ps followed by a slower decay which strongly varies with temperature. If not noted otherwise, all amplitudes and lifetimes in this paper refer to the latter decay.
 
The overall temporal evolution of $\Theta_\text{K}$ and its temperature dependence up to \unit{40}{K} are qualitatively similar to many other studies on TMDs, i.e. there are ns-scale spin lifetimes at low temperatures which strongly decrease towards higher temperatures.\cite{NatureComm.6.896, PhysRevB.90.161302, NanoLett.16.5010, NatPhys.11.830, NanoLett.15.8250,Kim2017Jul,Goryca2019Mar} It is important to note that even the reported $\mu$s spin lifetimes in TMDs can only be measured at cryogenic temperatures and undergo the same strong decay, yielding very short lifetimes at elevated temperatures.\cite{PhysRevLett.119.137401, Science.360.893, Kim2017Jul} In contrast, we observe strikingly different spin dynamics at higher temperatures: at \unit{40}{K} (green curve in Fig.~1(a)) $\Theta_\text{K}$ approaches zero for delay times $\Delta t>\unit{4}{ns}$, which results from the strong decrease of the spin lifetime. But at higher temperatures an additional spin signal with negative sign emerges. It first appears as a small signal with negative sign for time delays larger than \unit{1}{ns} at \unit{60}{K} (orange curve), and becomes fully developed at \unit{100}{K} at all $\Delta t$. Remarkably, there is a very weak temporal decay of $\Theta_\text{K}$, indicating extremely long spin lifetimes. Most interestingly, this trend holds over the whole temperature range up to RT, where we observe spin lifetimes of up to \unit{100}{ns}, 2 orders of magnitude larger than previously reported values at this temperature. We note, however, that the measurements at RT underwent temporal changes on laboratory time scales, which result in a spread of lifetimes ranging from \unit{24}{ns} up to \unit{100}{ns} for subsequent measurements taken over several days (see Fig.~1(b), where we plot $|\Theta_\text{K}|$ on a semi-logarithmic scale for easier comparison). We emphasize that the spin lifetimes did not decrease continuously with each successive measurement, but fluctuated without a discernible pattern (measurements in Fig.~1(b) are labeled in chronological order). Such temporal changes in the spin properties of TMD flakes were already reported previously\cite{PhysRevB.95.235408, APL.111.082404} and will be discussed further below in more detail.

\begin{figure*}[tb]
		\includegraphics[width=\linewidth]{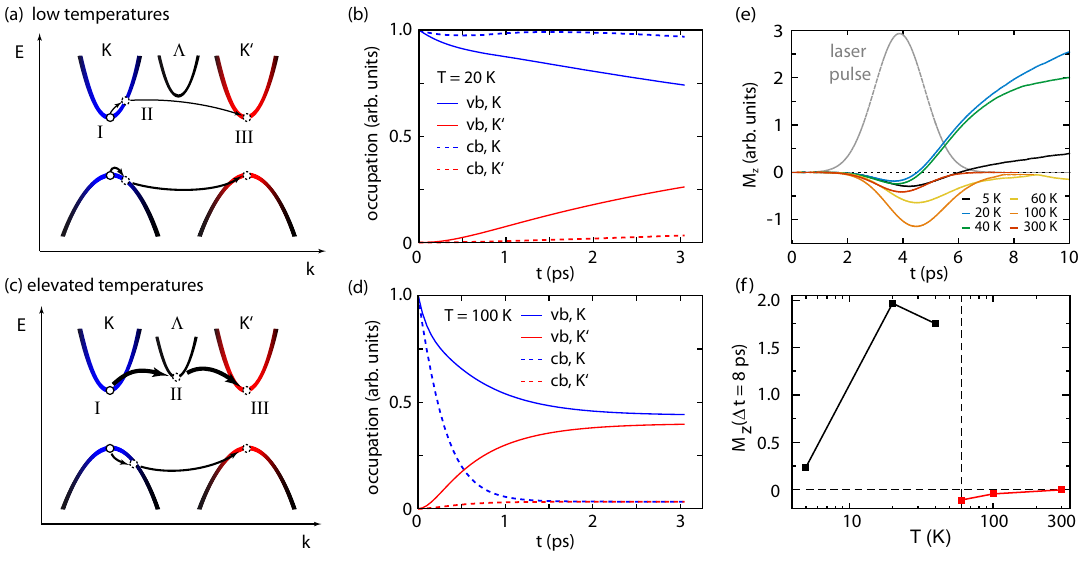}
	\caption{(a) and (c) - Representation of the bands' spin-texture (blue for spin-down, red for spin-up, and black for mixed states) near K and K' at low (a) and elevated (c) temperatures, illustrating the phonon-mediated allowed transitions. White dots mark the electron-hole pair's position in the band structure. At higher temperatures, electrons can transit to an intermediate valley at $\gL$. (b) and (d) - time-dependent evolution of the integrated occupations of both valence band (vb) and conduction band (cb) in the K and K' valleys at \unit{20}{K} (b) and \unit{100}{K} (d) for a prepared ideal exciton. (e) Time evolution of the magnetization at different temperatures when a real pulsed laser field and the e-ph interaction are both turned on. (f) Magnetization at a time delay of $\Delta t = \unit{8}{ps}$ as a function of temperature, showing the initial increase in amplitude and the crossover to a magnetization with reversed sign at around \unit{60}{K}, matching the experimental data in Fig.~1(c).}
	\label{figure2}
\end{figure*}

We note that the amplitudes of the TRKR traces in Fig.~1(a) are normalized in such a way that the discussed overall change in spin lifetimes is most clearly seen. The actual amplitudes are depicted in Fig.~1(c), where we plot the fitted Kerr amplitude $A_2$ of the longest spin signal seen in the bi-exponential fit at each temperature. Hence, for temperatures below \unit{60}{K} the black data points represent the spin states with a lifetime which decreases strongly with temperature, whereas for temperatures of \unit{60}{K} and above, the red data points correspond to the long-lived spin states with negative Kerr amplitudes. The trend of the Kerr amplitude shows a second intriguing feature besides the extraordinary long RT lifetimes: an increase of the Kerr rotation amplitude from \unit{10}{K} to \unit{50}{K}, followed by a sign reversal at the temperature at which the long-lived RT spin states emerge.

Before we explore the spin dynamics of other samples, we first focus on the theoretical understanding of the intriguing temperature dependence of the Kerr rotation amplitude in MoSe$_2$. The dependence can be understood essentially by changing electron-phonon scattering rates in the conduction and valence bands. We consider intrinsic MoSe$_2$, and take into account electron-hole pair (exciton) formation by the absorption of a pulsed laser field. We compute the temporal evolution of both conduction and valence band polarizations, solely driven by electron-phonon (e-ph) interactions, by performing real-time calculations with the \Yambo{} code~\cite{Sangalli2019} (see SI for details). These are based on approximations to the full Baym-Kadanoff equations.~\cite{pedro-paper,m.2012,PhysRevB.84.245110,PhysRevB.92.205304} In the simulations we use a laser pulse field with parameters (energy, temporal width, and intensity) which track those used in the experiment.

We find a striking resemblance between the temperature-dependence of the total MoSe$_2$ monolayer magnetization after the pulse and the measured Kerr amplitude. At \unit{60}{K} (see Fig.~2(e) and 2(f)) there is a complete inversion of the magnetization signal, consistent with the inversion of the Kerr signal shown in Fig.~\ref{figure1}(a). This behaviour can be explained by the differences in the dynamics of carriers in the conduction band at different temperatures (see SI for an extensive analysis). 
The underlying physical processes are illustrated in Fig.~\ref{figure2} (a). The direct transition between the states at the K and K' points is forbidden by selection rules. However, only the states at band extrema are purely spin-up or spin-down. Those at neighboring wave-vectors contain a mixture of both spins. As a result, inter-valley scattering of a pure spin state requires a two-step electron-phonon process: first, by absorbing an acoustic phonon, carriers are sent away from the band extrema to nearby states (transition from I to II in Fig.~\ref{figure2} (a)); then, temperature permitting, a phonon with a large wave-number can be absorbed or emitted, and the carrier can be transferred to the other valley (transition from II to III in Fig.~\ref{figure2} (a)). 

To separate the effects arising from e-ph interaction from those created by the laser field, we first discuss the case of a simple initial state, where we prepare an electron-hole pair exactly at the band extrema in the K valley (further details in the SI). Both carriers are independently allowed to interact with the phonon bath, and we track the time evolution of carrier populations in each band. These are shown in Fig.~\ref{figure2} (b) and (d) where we depict the temporal changes of both conduction (electrons, shown as the dashed lines) and valence band occupations (holes, solid lines). At low temperatures (Fig.~\ref{figure2} (b)) holes predominantly scatter into the other valley, leading to faster depolarization of the valence bands. The overall magnetization predominantly comes from the electron population imbalance, which barely reduces over the 3 ps time scale. 

The situation changes qualitatively at higher temperatures (Fig.~\ref{figure2}(d)), where the e-ph scattering rates of the conduction band states have drastically increased. This is largely due to an additional scattering channel through a valley at $\gL$ (also see supplementary Fig.~S4) which becomes accessible by thermal excitation (transition from I to II in Fig.~\ref{figure2}(c)). This new scattering channel turns out to be faster than those available to the hole population. Electrons will thus depolarize more quickly, leaving behind an imbalanced distribution of holes, which magnetizes the system, but now with a reversed sign. We point out that the valley populations shown in Figs.~\ref{figure2}(b) and (d) are obtained by summing all state occupations for a given band near either the K or K' points. As temperature increases, electrons and holes can spread more widely throughout the Brillouin zone, so the total number of electrons and holes inside the region of integration need not to add up to 1. 

The final step is the full simulation with both e-ph interaction for different temperatures and the optical excitation by the laser pulse. The resulting time-evolving magnetization is shown in Fig.~\ref{figure2}(e).The magnetization is evaluated by summing up over all occupied states in the whole Brillouin zone, taking into account the different spins of electrons and holes. This should not be confused with the valley occupations shown in Figs.~\ref{figure2} (b) and (d).
Initially the system has no magnetization, as the laser field creates an equal population of electrons and holes. Only after carriers start to spread through the Brillouin zone, due to e-ph interaction, do we observe an magnetization due to the phonon-induced spin flip processes. At low temperatures ($T=5$~K), since holes scatter faster than electrons, the density of electrons is spin polarized ($M_z>0$ at 10~ps). Increasing the temperature strongly increases $M_z$ up to 20~K, before it drops and reverses its sign at 60~K ($M_z<0$ at 10~ps). This behavior nicely matches the Kerr rotation amplitudes in Fig.~\ref{figure1}~(c) (see also Fig.~2(f)) and results from the difference in thermal activation of the e-ph scattering rates between electrons and holes, demonstrating that the e-ph interaction is the key to understand the initial formation of the valley polarization in the TRKR measurements.

\begin{figure*}[tb]
		\includegraphics[width=\linewidth]{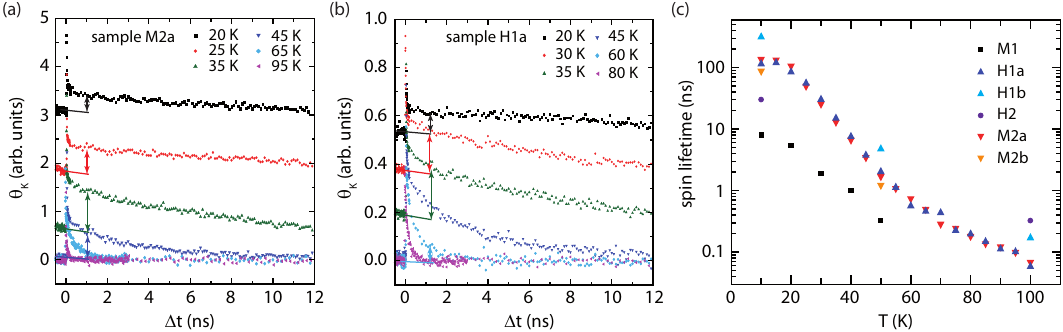}
	\caption{(a) and (b): Temperature dependent TRKR traces for MoSe$_2$ samples M2a and H1a, respectively. The overall Kerr amplitude decreases as there is less remaining valley polarization from previous pump pulses (see $\Theta_K$ signal for time delays $\Delta t < \unit{0}{ns}$) at elevated temperatures due to the decrease in spin lifetime. In contrast, the Kerr amplitude induced by the pump pulse at $\Delta t = \unit{0}{ns}$ (see arrows) first increases with increasing temperature before it drops significantly at higher temperatures. (c) Spin lifetime of the long-lived spin signal from the bi-exponential fit ($\tau_{s,2}$) of all MoSe$_2$ samples vs. temperature. For sample M1 we only show low temperature spin lifetimes, which interestingly are the shortest of all samples.}
	\label{figure3}
\end{figure*}

To support these findings, we next discuss the temperature dependent spin dynamics on five other MoSe$_2$ samples. The capital letter of the sample’s name indicates the supplier (``M'' for Manchester Nanomaterials and ``H'' for HQ Graphene), followed by the sample number indicating the crystal used for fabrication. A final small letter - if present - represents measurements done at different stages of the fabrication process (see below). For all samples we observe the temperature activated increase in valley polarization, which can be seen in the TRKR curves for samples M2a and H1a in Figs.~3(a) and 3(b), respectively. At first glance, there is a continuous decrease of the Kerr rotation amplitude with increasing temperature. In this context, however, it is important to note that, in contrast to sample M1 (Fig.~1(a)), the other samples have spin lifetimes of more than \unit{100}{ns} at \unit{10}{K} (Fig.~3(c)), which is much longer than the laser repetition interval of \unit{12.5}{ns}. Therefore, the Kerr rotation signal from previous pump pulses is not fully decayed prior to the arrival of the pump pulse at $\Delta t=\unit{0}{ns}$ (see non-vanishing $\Theta_K$ for time delays $\Delta t < \unit{0}{ns}$ in Figs.~3(a) and 3(b)). As a result, the constructive superposition of the valley polarization from successive pulses increases the overall Kerr amplitude. As the spin lifetime decreases with temperature (see Fig.~3(c)), the residual valley polarization from previous pump pulses also drops significantly, which explains the overall decrease in the Kerr amplitude with temperature as seen in Figs.~3(a) and 3(b). 

We now focus on the additional valley polarization created by the pump pulses at $\Delta t = \unit{0}{ns}$, which is highlighted by arrows in Figs.~3(a) and 3(b). The gain of the Kerr rotation amplitude first increases with temperature, then it strongly decreases at around \unit{60}{K} (see light blue curves) for all samples. This temperature corresponds very well to the crossover predicted by our simulation in Fig.~2(f) and to the data of sample M1 in Fig.~1(c), demonstrating that the temperature-dependent increase of the valley polarization probed by the Kerr effect is a hallmark for the spin dynamics in MoSe$_2$. We note, however, that we do not observe the sign reversal of the Kerr signal at higher temperatures, or the long lived RT spin signal, in any other sample. In contrast, the spin lifetime in those samples dramatically decreases at high temperatures, reaching values on the order of \unit{100}{ps} at \unit{100}{K} (Fig.~3(c)).

To unravel why only sample M1 shows the intriguing high temperature spin dynamics, we conduct photoluminescence (PL) measurements on all samples, as summarized in Fig.~4. For PL excitation we use a continuous wave (cw)-laser with an energy of \unit{2.33}{eV}, at a similar laser power and spot size as used for TRKR measurements. The spectra in Figs.~4(a) to 4(c) show typical emission from neutral ($X_0$) and from charged excitons ($X_{+/-}$) which were fitted by Voigt functions (dotted lines in Fig.~4(a)) to extract both their width (Figs.~4(d)) and energy position (Figs.~4(e)).

Details in the sample fabrication process have a strong impact on the excitonic properties of the samples, as seen for samples H1a and M2a in Figs.~4(b) and 4(c). These two samples were fabricated from two different crystals from different suppliers, but the samples were fabricated simultaneously under identical conditions, i.e. using the same exfoliation method, the same chemicals for the same duration of time, and were stored and measured identically. Both samples show low-temperature spin lifetimes of over \unit{100}{ns}, which are, to our knowledge, the longest lifetimes reported so far for MoSe$_2$ at these temperatures (see Fig.~3(c)). Interestingly, the absolute values of the spin lifetimes and their temperature dependence are almost identical (Fig.~3(c), blue and red triangles) although their PL spectra differ qualitatively: the intensity ratio between neutral and charged exciton emission inverts, and a low energy tail in the spectra of sample H1a is observed. 

The impact of further post processing steps on the samples' properties is also seen in Figs.~4(b) and 4(c). For sample H1a, we used a hot pick up process, which is known for its surface cleaning effect,\cite{arXiv180300912P} to deposit a thin hBN flake on top of the MoSe$_2$ monolayer. As a result, sample H1b shows a significant reduction in the PL line widths and the vanishing of the low-energy tail (Fig.~4(b)). Independent of temperature, this encapsulation yields an overall increase of the spin lifetimes by a factor of three (Fig.~3(c), dark blue and light blue triangles). In contrast, sample M2b, which was made from sample M2a by annealing under ambient conditions at $\unit{180}{\degree C}$, 
shows additional defects as seen by the appearance of a low-energy tail 
(Fig.~4(c)). Simultaneously, we observe a decrease in measured spin lifetime after annealing (compare red and orange triangles in Fig.~3(c)). Overall, these results show that, as could be expected, samples with less disorder (e.g. defects or adsorbates) show longer spin lifetimes in TRKR measurements at low temperatures. We have seen that both the spin lifetimes and the excitonic features strongly depend on details in the fabrication process, indicating that the electronic properties are dominated by extrinsic effects. This is a probable explanation for the significant device-to-device variations found in literature.

\begin{figure*}[tb]
		\includegraphics[width=\linewidth]{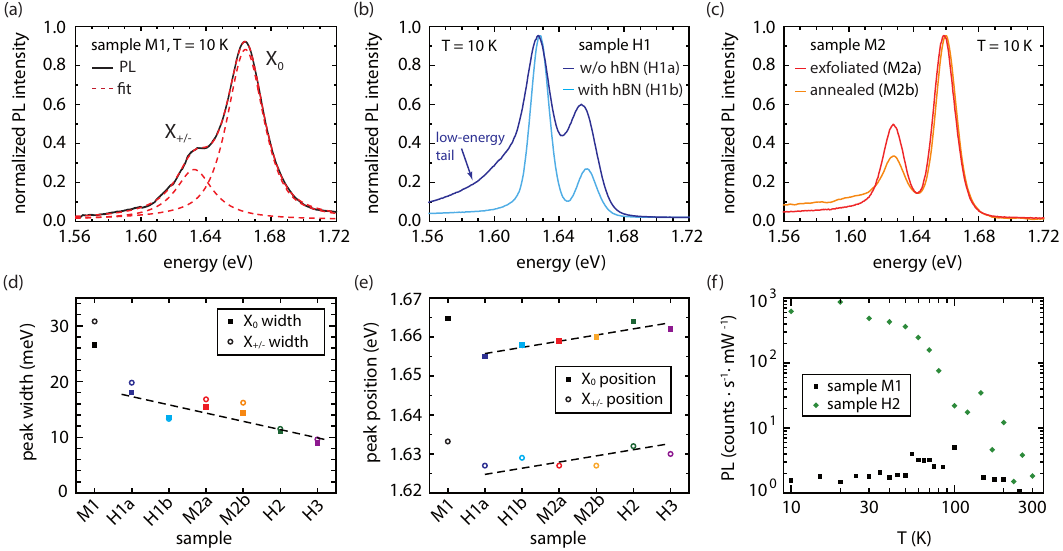}
	\caption{(a)-(c): Photoluminescence spectra of different samples measured at \unit{10}{K}. The spectra were fitted with Voigt-functions (dashed lines in (a)) to determine the peak positions and widths of both the neutral exciton ($X_0$) and the charged trion ($X_{+/-}$). The corresponding fit results for all of our samples at \unit{10}{K} are shown in (d) and (e), respectively. Here, it can be seen that the sample with the long-lived room temperature spin signal (sample M1) has by far the largest peak width, but at the same time exhibits the most blue-shifted peak positions (dashed lines are guides to the eyes). Sample M1 also distinguishes itself from other samples by showing a pronounced suppression of quantum yield over the whole temperature range, as seen in (f) where the normalized PL intensity of the neutral exciton peak is depicted for samples M1 and H2.}
	\label{figure4}
\end{figure*}

Sample M1 also follows the trend between disorder seen in PL and measured spin lifetime in TRKR: On the one hand, the low temperature spin lifetime is the lowest measured for all our samples (Fig.~3(c)), on the other hand, this sample shows by far the largest PL peak widths (Fig.~4(d)). These large peak widths are an indication of a significant variability in the potential landscape of the flake, induced either by disorder in the crystal lattice or by impurities.\cite{NatureComm.6.8315,2DMaterials.4.031011,PhysRevX.7.021026,arXiv180500127} However, there is also a striking difference in sample M1, which becomes obvious when comparing Figs.~4(d) and 4(e). For the other samples a narrower peak width is accompanied by a larger blue shift of the exciton peak position (see dashed lines, which are guides to the eyes). Despite the significant peak width of sample M1, its exciton peak position is the most blue-shifted among all samples, by \unit{1.665}{eV} which is higher than the vast majority of values reported in literature.\cite{Supplement} It was shown in several studies that either doping,\cite{NatureNanotechnology.12.856,NaturePhysics.13.255,NatNano.8.634} strain,\cite{Nanoscale.8.2589,PRB.88.121301,NanoLetters.13.3626} or the change of the dielectric environment by encapsulation with hBN\cite{ScientificReports.7.12383, 2DMaterials.4.031011,Supplement} can be associated with a red-shift of the exciton peak. As no sample except for H1b is in contact with hBN, and all samples are put on Si/SiO$_2$ substrates, we exclude the dielectric environment as a reason for the large blue-shift: the high exciton energy of sample M1 is therefore indicative of low overall doping and strain.

The large PL line widths of sample M1 indicate the presence of defects, but these do not carrier-dope the sample, which is apparent in the blue-shifted exciton peak and also the fact that the neutral exciton is much more pronounced than the trion in Fig. 4(a). The impact of defects on the electronic structure of sample M1 can also be seen in an unconventional temperature dependent quantum yield, shown in Fig.~4(f) where we normalized the total counts of the exciton peak by the integration time of the spectrometer and the power of the excitation laser. For comparison, sample H2 shows the expected temperature-dependent decrease in quantum yield for MoSe$_2$, which is attributed to the thermal activation of dark excitons.\cite{NatureCommunications.6.10110,arXiv180500127} On the other hand, the quantum yield of sample M1 is orders of magnitude lower at cryogenic temperatures and shows only a negligible temperature dependence. This behavior can be explained by defects which provide non-radiative recombination channels for the photo-excited charge carriers.\cite{arXiv180500127}

\begin{figure}[tb]
		\includegraphics{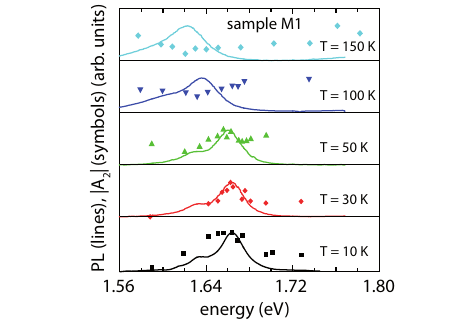}
	\caption{Comparison between PL spectra and amplitude $|A_2|$ of the long-lived Kerr rotation signal at varying pump energies for different temperatures. Whereas the pump scans follow the PL spectra for low temperatures, the excitonic energy resonance in the Kerr measurements vanishes for higher temperatures. Each horizontal line is the zero baseline for both the pump scan and the PL spectrum right above it.}
	\label{figure5}
\end{figure}

To explore the origin of the extraordinary high-temperature spin dynamics in sample M1, we conducted energy-dependent Kerr rotation measurements, where   the pump energy was scanned across the exciton energy range at different temperatures (see Fig.~5). The Kerr amplitude of the nanosecond spin signal (symbols) follows the PL spectra (solid lines) only at low temperatures. This resonant behavior weakens with increasing temperatures, and finally vanishes for $T>\unit{50}{K}$. At this point, the Kerr rotation amplitudes correspond to the high temperature long-lived spin signal with inverted sign (red data points in Fig.~1(c)). At temperatures below \unit{50}{K} the spin dynamics in sample M1 seems to be directly connected to the excitons. The fact that the pump scans at higher temperatures are almost energy independent indicates that the spin states are no longer hosted by the excitons. Rather, this indicates that there has been a transfer of spin polarization from excitons into other long-lived states. 

By combining all our findings, a complete picture of the spin dynamics observed in sample M1 can be obtained. The high-energy exciton peak position (Fig.~4(e)) indicates an overall small combination of strain and doping. This justifies our simulations of the temperature-dependent spin dynamics, where we assumed an intrinsic, undoped monolayer of MoSe$_2$ (Fig.~2). 
On the one hand, our simulations can explain the intriguing temperature-dependent Kerr amplitude of sample M1 seen in Fig.~1(c), which is due to the change of the ratio between phonon-induced conduction and valance band scattering rates. 
But on the other hand, our simulations predict that the magnetization drops to zero within a few ps at \unit{300}{K} (Fig.~2(e)). At lower temperatures we expect that the strong e-ph coupling in TMDs\cite{NanoLetters.17.4549, PRL.119.187402, NatureComm.8.14670, npj2DMaterialsandApplications.1.33, PRB.98.035302} will lead to a complete depolarization of the magnetization of itinerant states, both in the conduction and valence band, somewhere in the ps range. This was also the order of magnitude found in WSe$_2$.\cite{NanoLetters.17.4549} All these time scales are much shorter than the measured ns spin lifetimes, whether at high or low T.

The solution to this apparent discrepancy is that the strong electron-phonon scattering mechanism depicted in Fig.~2 only serves to prepare an initial spin state on the ps time scale. Accordingly, part of the initial magnetization must scatter into spin states which are protected over a ns time scale from the strong electron-phonon interaction of itinerant band states. In this respect it is important that excitons can be localized by trapping.\cite{PhysRevB.94.165301, PhysRevB.96.121404} The capture time by defects is measured to be around \unit{1}{ps},\cite{npj2DMaterialsandApplications.1.15} which is well within the lifetime of our simulated magnetization process. Furthermore, defect-bound excitons in monolayer TMDs are found to exhibit photoluminescence lifetimes in the hundreds of ns range.\cite{PhysRevB.96.121404, PhysRevLett.121.057403} However, this picture only holds for low temperatures: for temperatures approaching \unit{100}{K} the thermal energy exceeds the typical localization energy of excitons in such traps.\cite{PhysRevB.94.165301, APL.105.101901} This explains the  strong decrease of spin lifetimes with temperature up to \unit{100}{K}.

Another possible explanation for the ns lifetimes at low temperatures might be connected to polarons, which recently have drawn quite some attention in TMDs.\cite{NaturePhysics.13.255,PhysRevLett.114.146404,PRL.119.187402,PhysRevB.95.035417,JournalofAppliedPhysics.123.204308,JournalofAppliedPhysics.123.214303} We think that it is reasonable to assume that (dark) excitons\cite{PhysicalReviewMaterials.2.014002} can be protected against the strong electron-phonon interaction through a polaronic reconstruction of the combined electron and phonon system, which goes beyond the theory implemented in our real time simulations. Both explanations (trapped excitons and polaron-exciton complexes) depend on excitons and therefore would explain why the Kerr amplitude at low temperatures follows the excitonic features in the PL spectra of Fig.~5.

The long-lived spin states at higher temperatures, however, do not scale with the number of excitons created, as seen in Fig.~5, and therefore must have another origin. We first note that due to different inter-valley scattering rates of electrons and holes, their polarization can be transferred to resident charge carriers as soon as the excitons recombine.\cite{NatureComm.6.896} 
As explained above, we do not expect that these resident charge carriers exhibiting ns spin lifetimes are itinerant states of either the conduction or the valence band. Instead, a recent combined scanning tunneling spectroscopy and photoluminescence study identified defect states in the band gap of monolayer MoSe$_2$ samples which show a suppressed quantum yield similar to our sample M1 in Fig.~4(f).\cite{arXiv180500127} The large peak widths of the PL spectra in sample M1 are also an indication of large spatial fluctuations of the electrostatic potential, which could create puddles of hole or electron doping. Because of the overall small doping of sample M1, these puddles may be well separated. We imagine these puddles to be similar to those in graphene,\cite{NaturePhysics.4.144} and in fact first signatures of localized charge puddles have been seen in $\mu$PL measurements of TMD materials.\cite{ACSNano.11.2115} Defect states in these charge puddles would result in strong localization, which may protect the spin from phonon-induced scattering. Such a picture can also explain the strong variations in the repeated room temperature measurements in Fig.~1(b), as the constant desorption and absorption of volatile gas species at RT constantly changes the charge puddles.

In conclusion, we have demonstrated that electron-phonon interaction is the key for the ultrafast formation of a valley polarization in MoSe$_2$ after optical excitation by circularly polarized light. Different phonon-induced scattering rates of both valence and conduction band states were shown to result in a temperature dependent reversal of the valley polarization in good agreement to measurements from time-resolved Kerr rotation. While electron-phonon scattering involving itinerant band states is expected to result in a rapid depolarization, the measured spin lifetimes of up to 100 ns both at low temperatures and at room temperature indicate that the spin polarization scatters into spin states which are protected from intrinsic electron-phonon interaction.

\begin{acknowledgements} 
This project has received funding from the European Union's Horizon 2020 research and innovation programme under grant agreement No 785219. Supported by the Helmholtz Nanoelectronic Facility (HNF)\cite{HNF} at the Forschungszentrum J\"ulich.
PMMCM and MJV acknowledge funding by the Belgian FNRS (PDR G.A. T.1077.15-1/7), and the Communaut\'e Fran\c{c}aise de Belgique (ARC AIMED G.A. 15/19-09). 
ZZ acknowledges financial support by the Ramon y Cajal program (RYC-2016-19344), the CERCA programme of the Generalitat de Catalunya (grant 2017SGR1506), and by the Severo Ochoa programme (MINECO, SEV-2017-0706), and the EC 
H2020-INFRAEDI-2018-2020 MaX "Materials Design at the Exascale" CoE (grant No. 824143).
Computational resources have been provided by the Consortium des Equipements de Calcul Intensif (CECI), funded by FRS-FNRS G.A. 2.5020.11; the Zenobe Tier-1 supercomputer funded by the Walloon Region under G.A. 1117545; and by a PRACE-3IP DECI grant 2DSpin on Beskow (G.A. 653838 of H2020).

The authors declare no competing financial interest.
\end{acknowledgements}

\end{document}


\title{Supplement: Spin States Protected from Intrinsic Electron-Phonon-Coupling Reaching 100 ns Lifetime at Room Temperature in MoSe$_2$}

\author{Manfred Ersfeld}
\affiliation{\aachenphys}
\author{Frank Volmer}
\affiliation{\aachenphys}
\author{Pedro Miguel M. C. de Melo}
\affiliation{\liege}
\affiliation{\etsf}
\author{Robin de Winter}
\affiliation{\aachenphys}
\author{Maximilian Heithoff}
\affiliation{\aachenphys}
\author{Zeila Zanolli}
\affiliation{\etsf}
\affiliation{\barcelona}
\affiliation{Institute for Theoretical Solid State Physics,
RWTH Aachen University, D-52056 Aachen, Germany}
\author{Christoph Stampfer}
\affiliation{\aachenphys}
\affiliation{\pgi}
\author{Matthieu J. Verstraete}
\affiliation{\liege}
\affiliation{\etsf}
\author{Bernd Beschoten}
\affiliation{\aachenphys}

\date{\today}

\maketitle

\section{Temporal changes in spin lifetime at room temperature}
We note that the room temperature TRKR signal of sample M1 seen in Fig. 1(b) of the main manuscript did not decrease continuously with each successive measurement, but rather decreased and increased without discernible pattern. As such kind of change in device properties may be explained by different kind of adsorbates on top of the flake, we tried to keep the vacuum conditions for each of our measurements as reproducible as possible. Unfortunately, the two-dimensional nature of the TMD flake demands technological requirements that are beyond the possibilities of our setup. In surface science it is a well-known fact that reproducible conditions can only be achieved under ultra-high vacuum conditions (below $\unit{10^{-9}}{mbar}$). In the experiment we used a continuous flow cryostat that can be pumped down at room temperature to a base pressure of  $\unit{10^{-6}}{mbar}$. From a surface science point of view, such conditions are “dirty” (it is equal to a monolayer formation time of contaminating gas species on top of a clean surface of only \unit{1}{s}). It is known that air-born hydrocarbons cover the surface of 2D materials once they were exposed to air.\cite{Carbon.61.33, Langmuir.31.8429} Even under vacuum conditions the binding energy of long-chained hydrocarbons leads to desorption temperatures above RT.\cite{Carbon.44.2931, Carbon.61.33} But as our cryostat cannot be baked-out at high temperatures, we do not have control over the residual gas composition, which e.g. condenses on the sample surface during a cool-down.

\begin{figure}[tb]
	\includegraphics{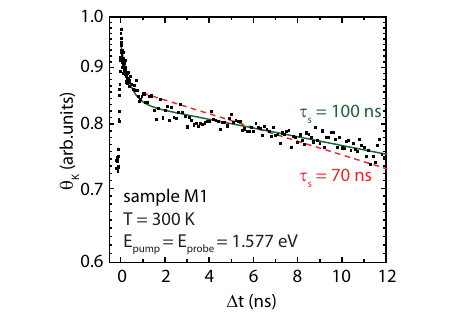}
\caption{Time resolved Kerr rotation curve of sample M1 which shows a decay time of around \unit{100}{ns} at room temperature. Fits were done by fixing the longer of two decay times of a two-exponential decay function to either \unit{70}{ns} or \unit{100}{ns}.}
\label{SupFig1}
\end{figure}

As the longest spin lifetimes exceed by far the adjustable delay time (which is limited by the laser repetition interval of \unit{12.5}{ns}), the question arises how reliable the fitted lifetimes are as soon as they are in the \unit{100}{ns} range. In Fig. S1 of this supplemental material we fitted one of the room temperature measurements on sample M1 with spin lifetimes on the order of \unit{100}{ns} by a two-exponential decay function and fixed the longer of the two spin components either to \unit{70}{ns} or \unit{100}{ns}, respectively. For an assumed spin lifetime of \unit{70}{ns} the majority of data points lie either below or above the dashed line for time delays lower or higher than $\Delta t = \unit{6}{ns}$, respectively. Because of this systematic behavior of the residuals, the actual lifetime has to be longer than \unit{70}{ns}. We repeated such fits with increasing but fixed values of the spin lifetime. At $\tau_s = \unit{100}{ns}$ the signal-to-noise ratio of the curve masked any possible systematic trend in the residuals. Hence, the true spin lifetime has to be at least \unit{100}{ns}.

\section{Challenges in the measurement of the RT spin signal}
We note that the Kerr rotation amplitude of the RT spin signal in sample M1 is only in the order of some $\mu$rad, which is significantly lower than the spin signal at lower temperatures, where the Kerr rotation reaches tens of $\mu$rad. However, most likely due to the 2-dimensional character of the TMD, even the amplitudes of the low-temperature spin signal fall short to typical Kerr rotation values measured e.g. in GaAs or ZnO where values of several hundreds of $\mu$rad or even mrad can be measured.\cite{PhysRevLett.105.246603, PSSB.251.1861}

There are two challenges with these small amplitudes: First, a good adjustment of all optical elements and long integration times are needed to obtain a decent signal-to-noise ratio. A single delay scan at RT (like it is shown in Fig.~2(b) of the main manuscript) takes at least one hour. Now it is important to note that changes in the spin signals predominantly occur at higher temperatures, whereas the spin properties are very stable at low temperatures.\cite{PhysRevB.95.235408} As mentioned above, this can be explained by adsorbates on the TMD flake which condense at cryogenic temperatures, but can freely adsorb and desorb at higher temperatures. The unstable spin properties at higher temperature combined with the required long acquisition times for a single delay scan does not allow to measure e.g. consistent Kerr resonance curves (i.e. probe scans). 

The other challenge considering the small Kerr rotation signal is a possible small background in the Kerr rotation data. A spurious electronic signal responsible for such a background can, e.g., be generated by the balanced diode bridge we use for our optical Kerr measurements.\cite{PhysRevB.56.7574} Because of the finite common mode rejection of the used differential amplifier in such an optical bridge, each balanced diode bridge will output a small spurious differential signal. Depending on the exact adjustment of the whole laser system (see schematic setup in the supplemental material of Ref.~\citenum{PhysRevB.95.235408}) and the used settings of lock-in and voltage amplifiers, we normally observe a background signal up to several hundreds of nrad. We note that such a small background signal can be easily overlooked if the actual spin signal is e.g. several hundreds of $\mu$rad large. However, in our case (in which the RT spin signal is only several $\mu$rad large) such a constant background can have a non-negligible influence on the determined spin lifetime. This is especially the case if the spin lifetimes exceed the measurable delay time.

If the signal was showing, e.g., spin precession or any other kind of magnetic field dependence, such a background could in principle be determined at high enough magnetic field strengths. However, our samples do not show any magnetic field dependence. Therefore, we determined the background signal by reference measurements in which we moved the sample approximately $\unit{100}{\mu m}$ away from the laser spots and measured the resulting signal on the Si/SiO$_2$ wafer. The determined background was subtracted from all subsequent measurements.

We note that the above mentioned spurious signals from the electronics can be seen as constant over the measurement time and therefore cannot explain the clear, reproducible jump at zero delay and the exponential decay in our Kerr data. Furthermore, we regularly checked if the signal, which we identified as the spin signal, changes sign by inverting the helicity of the pump laser. Additionally, we note that the spin signal vanishes if the pump beam is linearly polarized or if either the pump or probe beam is individually blocked. Therefore, we exclude other laser-induced effects as the origin of the long-lived signal.

\begin{figure}[tb]
	\includegraphics{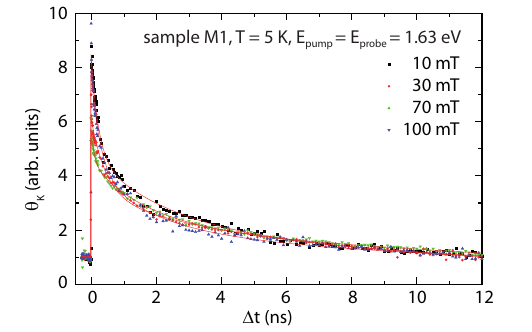}
\caption{TRKR curves of sample M1 in the trion regime at \unit{5}{K} for different magnetic fields applied in-plane to the MoSe$_2$ flake. Red curves are fits according to a multi-exponential fit function.}
\label{SupFig2}
\end{figure}

\section{Fitting of the Kerr rotation curve}

The Kerr rotation signal $\Theta_{K}(t)$ for all samples were fitted over the whole temperature range by a bi-exponential fit function of the form:
\begin{equation}
\Theta_{K}(t) =\sum_{i=1}^2 A_i \cdot \exp \left( -\frac{t}{\tau_\text{s,i}} \right),
\label{eq:Fit}
\end{equation}
where $A_{i}$ are the amplitudes and $\tau_\text{s,i}$ are the spin lifetimes for each component. As the absolute value of the Kerr rotation is very sensitive to the alignment of the laser system in general and the overlap of pump and probe beam in special, great care was taken for the temperature dependent measurements shown in Figs. 1(c), 3(a), and 3(b) of the main manuscript. For each temperature, we spatially mapped both Kerr signal and reflectivity over the whole flake to guarantee the probing of the same spot on the sample. This is important as thermal expansion of the cryostat changes the relative position of the sample in respect to the laser beams. Furthermore, for each temperature we checked that pump and probe beam were in good overlap to each other. Nevertheless, unavoidable slight variations in the probed location of the sample and minimal variations in the overlap cannot be completely avoided and are the most likely explanation for the scattering of the Kerr amplitude in Fig. 1(c) of the main manuscript.
  
\begin{figure*}[t]
	\includegraphics{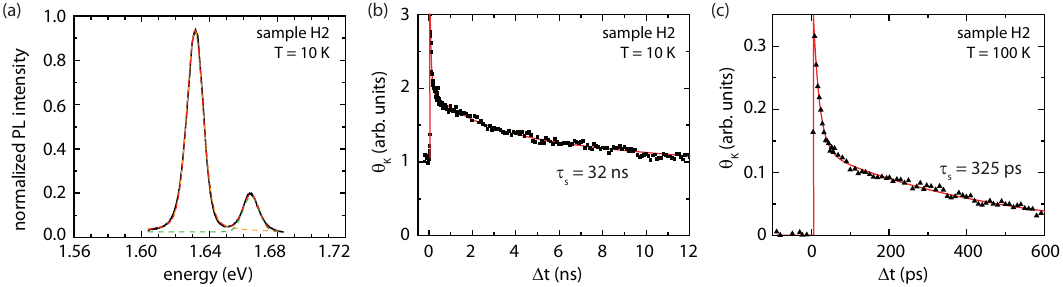}
\caption{(a) Photoluminescence spectra for sample H2 at $T=\unit{10}{K}$. The red dashed line is the sum of two Voigt functions fitted to the exciton and trion peak (green and orange dashed line, respectively). (b) and (c): Time-resolved Kerr rotation curves of sample H2 at $T=\unit{10}{K}$ and $T=\unit{100}{K}$, respectively. The red lines are fits to a bi-exponential function.}
\label{SupFig3}

\end{figure*}

\section{Magnetic field dependent measurements}
In Fig.~S2 of this supplemental material we show TRKR curves of sample M1 in the trion regime at \unit{5}{K} for different magnetic fields. The field was applied in-plane to the MoSe$_2$ flake and thus transverse to the spin direction of the optically generated spin states. We note that the cryostat slightly moves when applying a magnetic field. This is important as PL and Kerr-measurements already demonstrated that a TMD flake can have spatially varying optical properties.\cite{arXiv160203568B} This may explain the marginal change between the curves within the first few ns, i.e. in the short-lived spin signal of the multi-exponential fit. On the other hand, the long-lived spin signal, which is the main focus in our main manuscript, shows no changes as a function of applied magnetic field and, hence, the curves are in good agreement to each other for time delays larger than $\Delta t = \unit{5}{ns}$.

\section{PL and Kerr data for sample H2}
In Fig.~S3 we show both the photoluminescence data at \unit{10}{K} and the TRKR curves for both \unit{10}{K} and \unit{100}{K} for sample H2. The measurement conditions are identical to the one given in the main manuscript for the other samples. Similar to samples H1a and M2a in Figs.~3(a) and 3(b) of the main manuscript, the spin lifetime exponentially decays towards higher temperatures. For this sample, no long-lived spin state with inverted sign can be measured within the resolution of our setup.

\section{Peak positions of neutral excitons}

As mentioned in the main manuscript, the sample with the long-lived RT spin signal (sample M1) exhibits an exciton peak position at quite high energies at \unit{1.665}{eV}. This energy is higher than the vast majority of values reported in literature as can be seen in Tab.~\ref{SupTab1}. It was shown in several studies that either doping,\cite{NatureNanotechnology.12.856,NaturePhysics.13.255,NatNano.8.634} strain,\cite{Nanoscale.8.2589,PRB.88.121301,NanoLetters.13.3626} or the change of the dielectric environment by encapsulation with hBN\cite{ScientificReports.7.12383, 2DMaterials.4.031011} can be associated with a red-shift of the exciton peak. As no sample except for H1b is in contact with hBN, and all samples are put on Si/SiO$_2$ substrates, we exclude the dielectric environment as a reason for the large blue-shift: the high exciton energy of sample M1 is therefore indicative of low overall doping and strain.

\begin{table}[ht]
  \caption{Reported values in literature about the exciton peak position in photoluminescence measurements at low temperatures in case of MoSe$_2$ on different substrates}
\begin{tabular}{|c|l|c|}
\hline
	reference & sample/substrate & exciton energy (eV) \\ \hline
	\citenum{NaturePhysics.13.255} & hBN/MoSe$_2$/hBN & 1.630 \\ \hline
	\citenum{2DMaterials.4.031011}& hBN/MoSe$_2$/hBN/SAM & 1.640 \\ \hline
	\citenum{NatureNanotechnology.12.856} & hBN/MoSe$_2$/hBN & 1.642 \\ \hline
	\citenum{2DMaterials.3.045008} & MoSe$_2$ Ti doped/SiO$_2$ & 1.648 \\ \hline
	\citenum{2DMaterials.3.045008} & MoSe$_2$/h-BN/SiO$_2$ & 1.648 \\ \hline
	\citenum{2DMaterials.4.031011} & hBN/MoSe$_2$/hBN & 1.650 \\ \hline
	\citenum{ScientificReports.7.12383} & hBN/MoSe$_2$/hBN & 1.652 \\ \hline
	\citenum{ScientificReports.7.12383} & MoSe$_2$/hBN & 1.655 \\ \hline
	\citenum{ScientificReports.8.2389} & MoSe$_2$/SiO$_2$ & 1.655 \\ \hline
	\citenum{Nanotechnology.28.395702} & MoSe$_2$ & 1.655 \\ \hline
	\citenum{2DMaterials.4.031011} & MoSe$_2$/SiO$_2$ & 1.657 \\ \hline
	\citenum{2DMaterials.3.045008} & MoSe$_2$ undoped/SiO$_2$ & 1.658 \\ \hline
	\citenum{2DMaterials.3.045008} & MoSe$_2$/Au/SiO$_2$ & 1.658 \\ \hline
	\citenum{NatureComm.4.1474} & MoSe$_2$/SiO$_2$ & 1.658 \\ \hline
	\citenum{OptMaterExpress.6.2081} & MoSe$_2$/SiO$_2$ & 1.658 \\ \hline
	\citenum{2DMaterials.3.045008} & MoSe$_2$/SiO$_2$ & 1.660 \\ \hline
	\citenum{ScientificReports.7.12383} & MoSe$_2$/SiO$_2$ & 1.660 \\ \hline
	\citenum{2DMaterials.3.045008} & MoSe$_2$/SiO$_2$ & 1.662 \\ \hline
	\citenum{ACSNano.10.1454} & MoSe$_2$/sapphire & 1.665 \\ \hline
	\citenum{APL.106.112101} & MoSe$_2$/SiO$_2$ & 1.667 \\ \hline
\end{tabular}
\label{SupTab1}
\end{table}

\section{Ground state and band structure}
\label{dft_gw}
\begin{figure}[ht]
	\includegraphics[width=\columnwidth]{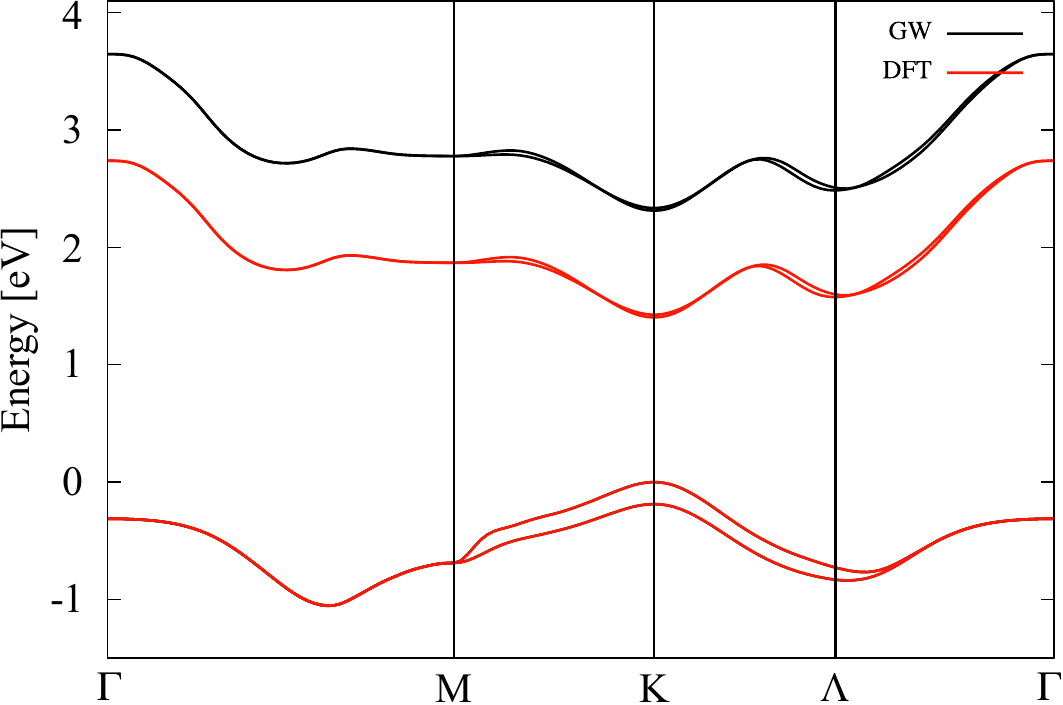}
\caption{DFT (red) and \gw (black) corrected bands for the \mose monolayer computed on a 30 by 30 k-point grid and plotted along the optical path. The \gw calculation changes the band gap at K from 1.40 eV to 2.34 eV. 
}
\label{fig:dft_gw}
\end{figure}

Electronic ground state properties were computed using Density Functional theory (DFT) with the fully-relativistic pseudo-potentials generated by the ONCVPSP code~\cite{PhysRevB.88.085117} and stored in the Pseudo-Dojo repository~\cite{VanSetten2018}. A lattice parameter of 3.28 \ag was used. Both the in-plane cell and the positions of the atoms were relaxed. The final distance between the Mo and Se atoms was 2.51 \ag and the angle defined by the Se-Mo-Se atoms was 81.84$^\circ$. The ground state was computed with a 24 by 24 k-point grid which was then used to compute the databases on a 30 by 30 mesh. A DFT band gap of 1.40 eV was obtained, with the lifting of the degeneracy in the top valence bands at K due to spin-orbit coupling being 0.19 eV. The energy split for the bottom conduction bands at K, also due to spin-orbit, was 0.02 eV. The resulting DFT band structure is shown in red in Fig.~\ref{fig:dft_gw}.

To correct the band gap we performed a \gw calculation using the \Yambo~code~\cite{Sangalli2019}. A total of 240 bands and a cutoff of 2000 mHa were used to compute the screening, while 200 bands were needed for the Green's function and a cutoff of 6 Ha was used for the exchange component of the self-energy. The result was a scissor operator of \unit{0.94}{eV} which was applied to the band structure, as shown in black in Fig.~\ref{fig:dft_gw}. For the remaining of this work, we will label the bands as v1, v2, c1, and c2, referring to the first and second valence bands, and the first and second conduction bands, respectively and in increasing energy. 

\section{Optical absorption}
\label{bse}
\begin{figure}[ht]
	\includegraphics[width=\columnwidth]{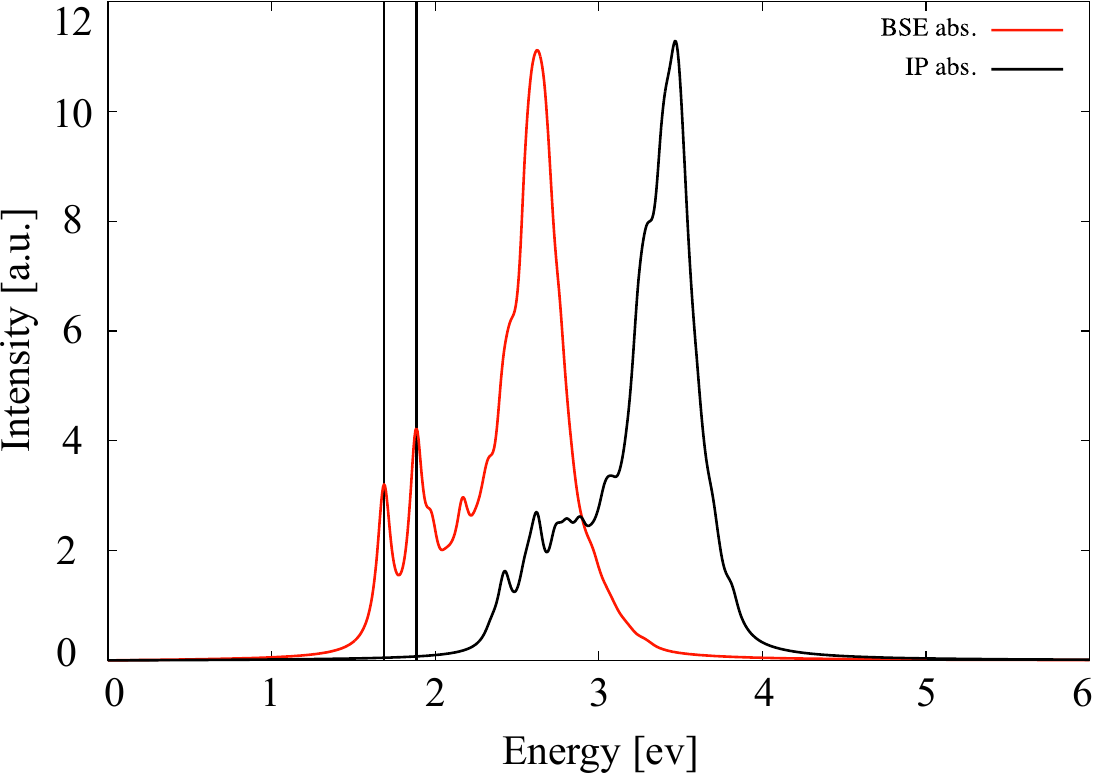}
\caption{\gw corrected (black) and BSE (red) absorption spectrum for the \mose monolayer computed on a 30 by 30 k-point grid. The vertical black lines indicate the positions of the first two excitons, A and B, at 1.69 eV and 1.90 eV respectively, consistent with the computed spin-orbit splitting. The binding energies for the A exciton is 0.66 eV and 0.44 eV for the B exciton.}
\label{fig:bse}
\end{figure}

To compute the absorption spectrum we need to go beyond \gw and include excitonic effects. This was also done with the \Yambo~code via the Bethe-Salpeter equation (BSE). The screening was computed within the COHSEX approximation and required a cutoff of 2000 mHa and 120 bands, while the exchange part needed a cutoff of 10 Ha. In the building of the kernel the top two valence and the bottom two conduction bands were used.
The resulting absorption spectrum is shown in Fig.~\ref{fig:bse}. There we can observe the first two main peaks at 1.69 eV and at 1.90 eV, respectively. The energy difference between the two peaks is 0.21 eV, which is consistent with the spin-orbit splitting of the bands at K. 

\section{Real-time dynamics}
\label{rt}
Our approach to real-time dynamics is based on the Non-equilibrium Green's function (NEGF) formalism and in solving an approximated form of the Baym-Kadanoff equations (BKE)~\cite{pedro-paper,m.2012,PhysRevB.84.245110,PhysRevB.92.205304} for the single-time density matrix $\gr(t)$. The BKE can be written as~\cite{pedro-paper,m.2012,PhysRevB.84.245110,PhysRevB.92.205304}
\begin{equation}
\partial_t \gr_i(t) = \partial_t\gr_i(t)|_\text{coh} + \partial_t\gr_i(t)|_\text{coll},
\label{eq:bke}
\end{equation}
where $\partial_t\gr_i(t)|_\text{coh/coll}$ refer to the coherent/collision terms of the dynamics, respectively. Here $\gr(t)$ is expressed on a Kohn-Sham basis for electron-hole pairs, so $i = \{n,n',\kk\}$, where $n$ and $n'$ are spinorial band indices and $\kk$ is the crystal momentum.
The coherent term in Eq.~\eqref{eq:bke} contains the interaction with the laser field and coherent correlation effects. It can be written in a compact form as

\begin{multline}
 \partial_t\gr_i(t)|_\text{coh} = \Delta\epsilon_i\gr_i(t) +\\ +\left[\Delta\Sigma^{Hxc}(t),\gr(t)\right]_i + \left[U(t),\gr(t)\right]_i.
\label{eq:coh}   
\end{multline}

Here, $\Delta\epsilon_i = \epsilon_{\nk} - \epsilon_{\npk}$ is the energy difference between the quasi-particle energies of the system at equilibrium. The second term on the right-hand-side, $\Delta\Sigma^{Hxc}$, accounts for the change in the Hartree potential and the exchange-correlation self-energy at a given time $t$, while the last term includes the interaction with the laser field. The latter is given by $U(t) = -\gO\bf E(t)\cdot\bf d_i$, where $\gO$ is the volume of the unit cell, $\bf E(t)$ the time-dependent electric field, and $\bf d_i$ the matrix element of the dipolar moment\footnote{The dipolar approximation is well justified in this situation, since the wavelength of the laser is much larger than the dimensions of the unit cell.}. While the external electric field drives the excitation of the carriers, the change in the Hartree and exchange-correlation self-energies accounts, in linear order, for the electron-hole interaction by including excitonic effects. 

Relaxation of carrier populations is driven by the $\partial_t\gr_i(t)|_\text{coll}$, where the important mechanisms for valley scattering are included. It can be written as
\begin{multline}
    \partial_t\gr_i(t)|_\text{coll} = \sum_I\left[-\gamma_i^{(e,I)}(t,T)\gr_i(t)+\right.\\
\left.+\gamma_i^{(h,I)}(t,T)\bar\gr_i(t)\right]
\label{eq:coll}
\end{multline}
where $\bar\gr_i = \delta_{n,n'} - \gr_i$. The sum runs over an index $I$, which accounts for all three possible scattering mechanisms: electron-phonon ($I = e-p$), electron-electron ($I=e-e$), and electron-photon ($I=e-\gamma$). The generalised non-equilibrium time-dependent lifetimes $\gamma_i$ contain the scattering information coming from the Feynman diagrams. A derivation of Eq.~\eqref{eq:bke} and the specific form of these non-equillibirum lifetimes can be found in Ref.~\onlinecite{pedro-paper}. Once the evolution in time of the density matrix is computed, we can proceed to the calculation of optical properties. This is done by extracting the time-dependent occupations of the electronic levels at a given time $t$, $f_{\nk}$, which are the diagonal elements of the density matrix, i.e., $f_{\nk}(t) = \gr_{\nnk}(t)$. The main reasoning behind this is that measurements are performed at a point in time in which the initial dephasing effects induced by the laser pump have vanished already~\cite{pedro-paper,m.2012,PhysRevB.84.245110,PhysRevB.92.205304}. As such, we can easily compute whichever quantity necessary if it can be expressed as a function of the time-dependent occupations and energy levels~\cite{pedro-paper,m.2012,PhysRevB.84.245110,PhysRevB.92.205304}.

\section{The electron-phonon scattering mechanism}
\label{laser_elph}
\begin{figure}[ht]
	\includegraphics[width=\columnwidth]{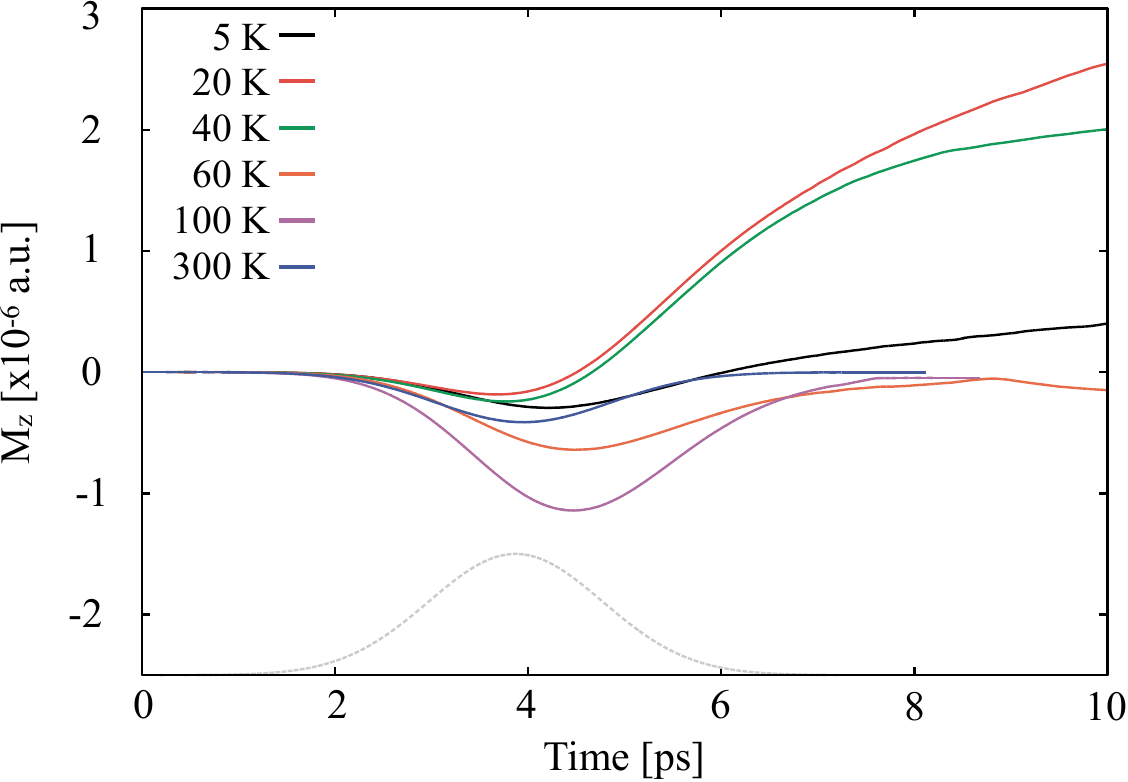}
\caption{Evolution in time of the magnetisation along $Oz$ of the \mose monolayer due to the laser pump, for 5 K (black), 20 K (red), 40 K (green), 60 K (orange), 100 K (purple), and 300 K (blue). Profile of laser intensity shown with grey dashed line.}
\label{fig:mag_time}
\end{figure}

We performed \ai  ~time-dependent simulations on a MoSe$_2$ monolayer using a right-hand circularly polarised laser tuned to the energy of the A exciton (1.69 eV) for temperatures of 5 K, 20 K, 40 K, 60 K, 100 K, and 300 K. The laser FWHM and intensity are set to the ones used in the experimental setup. In Fig.~\ref{fig:mag_time} we show the evolution in time of the $Oz$ component of the system's magnetisation ($M_z$) in time, for each temperature. 

\begin{figure}[ht]
\includegraphics[width=\columnwidth]{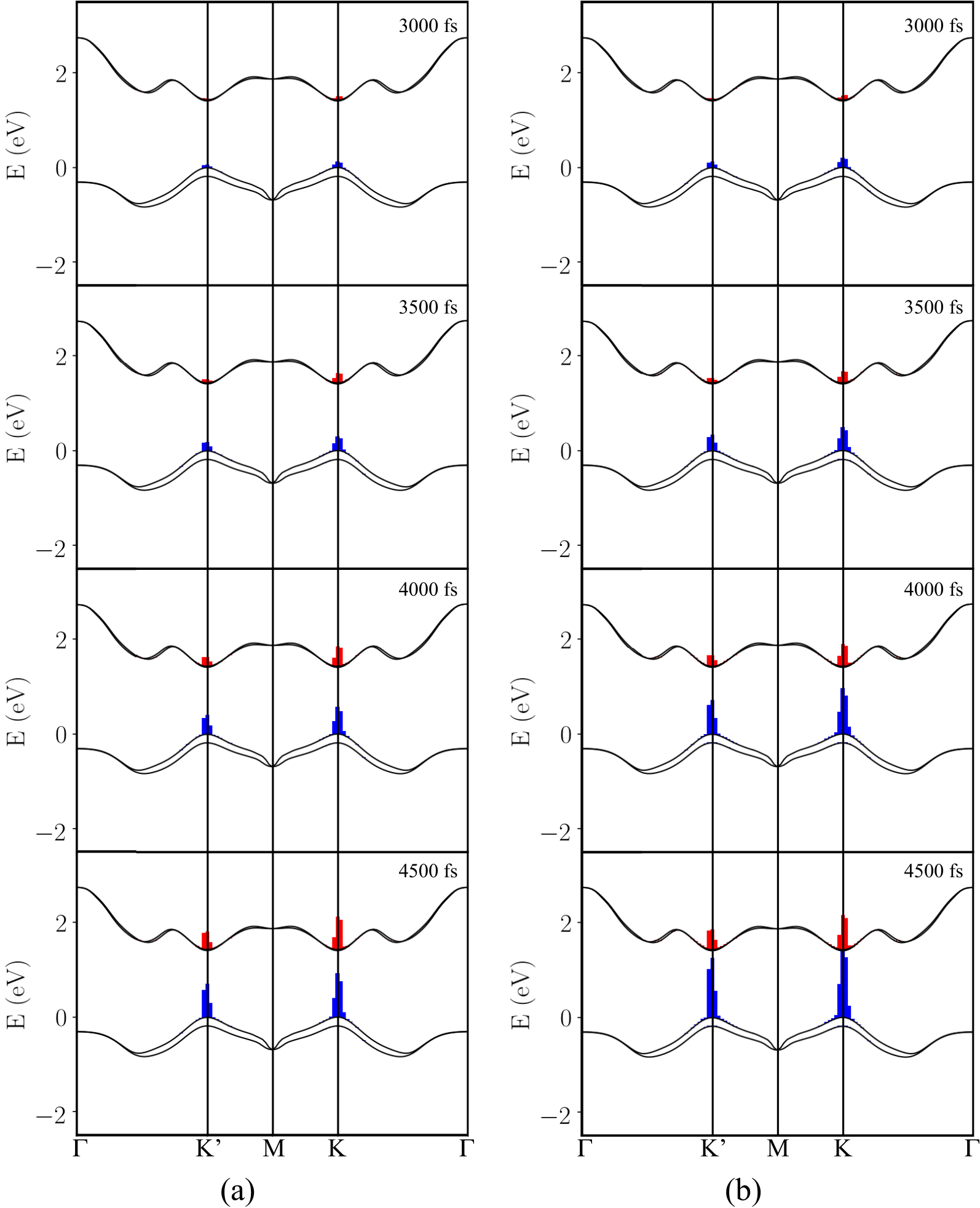}
\caption{Time snapshots of electron (red) and hole (blue) occupations at 40 K [(a)] and 60 K [(b)] under the effects of the external laser field.}
\label{fig:mag_time_1}
\end{figure}

There is a visible change in the time evolution of the magnetisation when temperature goes from 40 to 60 K, with $M_z$ no longer reversing signal in the latter. This magnetisation results from an imbalance between the electron and hole populations in the K and K' valleys. In Fig.~\ref{fig:mag_time_1} there are snapshots of the these populations in the band structure at 3 ps, 3.5 ps, 4 ps, and 4.5 ps for 40 K and 60 K, together with the change in time of the populations of the bands at each valley. There we see that the creation of pairs in the K valley (due to the laser pump) happens at a time scale close to that of the scattering of carriers towards the K' valley. This similarity in the time scales makes it difficult to understand how each process, excitation and scattering, contributes to overall change in behaviour of the magnetisation, so the two of them need to be picked apart. 

To do so we performed the two types of tests: one in which only the laser field was active and there was no electron-phonon scattering; and another where an electron-hole pair is created artificially and then allowed to interact with the phonon bath. 

\begin{figure}[ht]
	\includegraphics[width=\columnwidth]{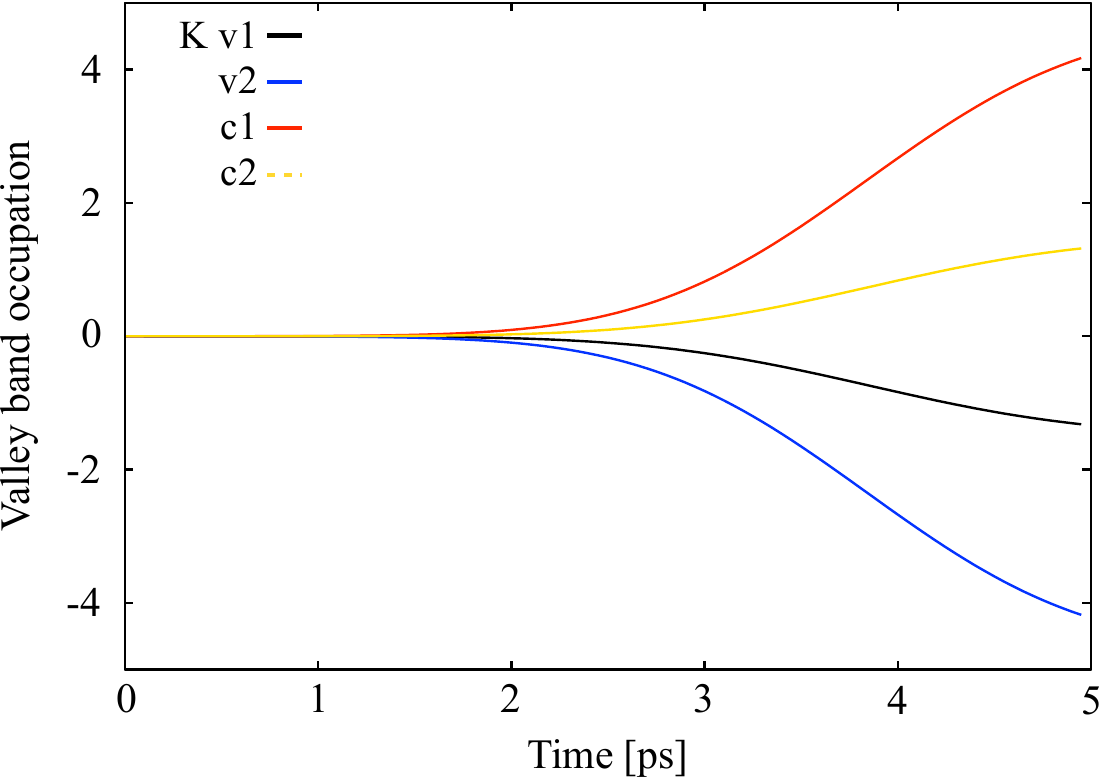}
\caption{Evolution in time of the valley band populations at the K valley due to the laser field only. Here hole populations are represented as negative so that there is no overlap between lines representing electron and hole populations.}
\label{fig:laser-20K}
\end{figure}

The effects of the laser pump acting alone on the \mose are represented in Fig.~\ref{fig:laser-20K}. As expected, the laser creates carriers only in the K valley, although not only bands v2 and c1 are populated, but also bands v1 and c2. Such effect has already been reported in literature~\cite{NanoLetters.17.4549}. Changes in the screening due to promotion of electrons to conduction states induce a renormalisation of the excitonic energies, such that the laser pulse is now capable of promoting electrons up from v1. Then spin and angular momentum selection rules ensure that these electrons are moved to the c2 band.

Taking into account these results we devised the second set of tests to study the effects of the electron-phonon dynamics. Here electrons are promoted ``by-hand'' from the valence to the conduction states at K exactly in three separate cases (which had to be consistent with the selection rules): from v2 to c1; from v1 to c2; and from v2 to c1 together with the promotion of an electron from v1 to c2. This way we could study how the phonons would affect the carriers created by the laser in a controlled and isolated fashion. 

\begin{figure}[ht]
\includegraphics[width=\columnwidth]{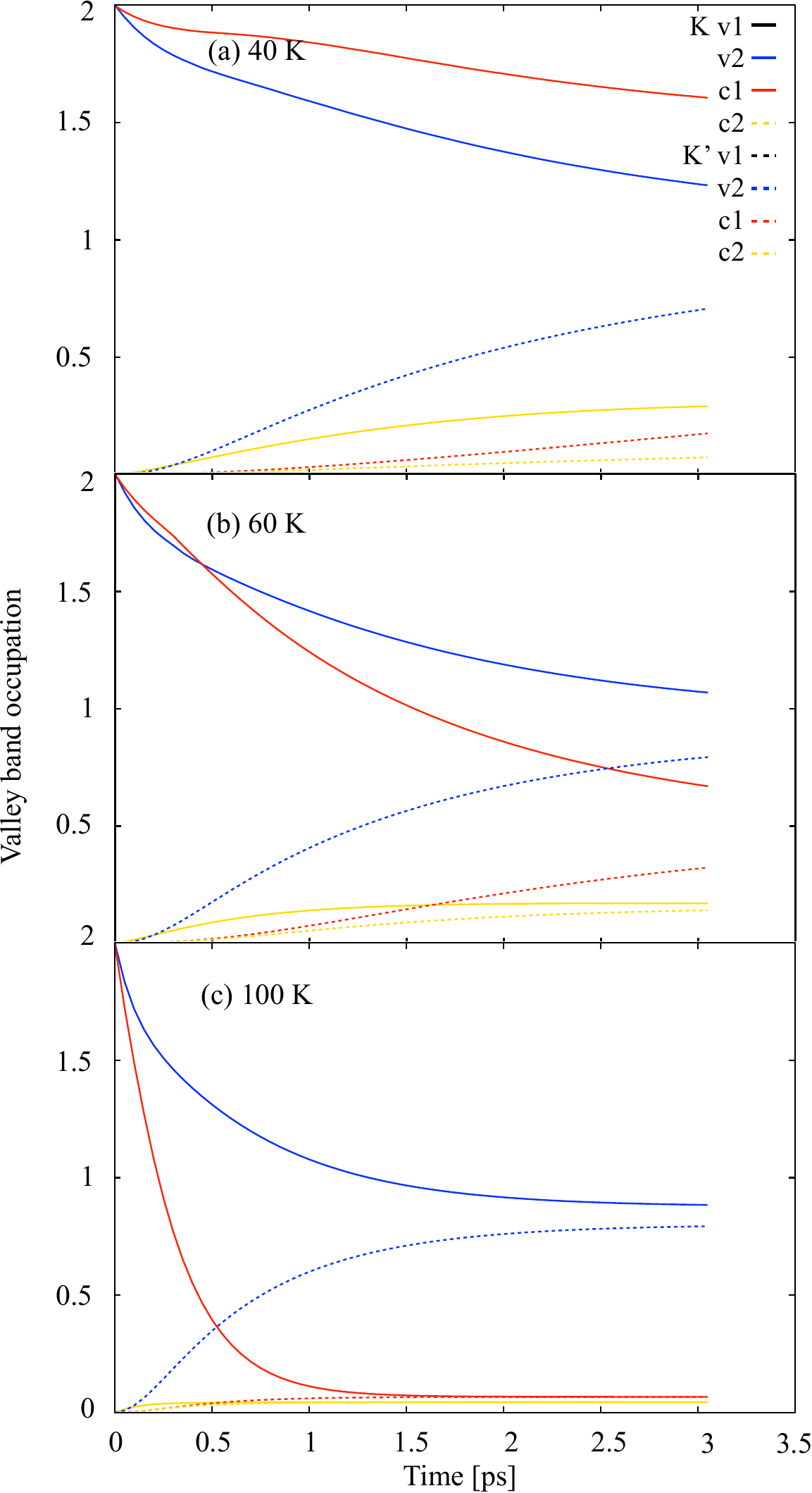}
\caption{Time-evolution of valley electron and hole populations for the artificial excited state in which electrons are moved from v2 to c1 at K, at 40K, 60K, and 100K. Full lines represent the populations in the K valley, while dashed lines show the populations of the corresponding band in the K' valley.}
\label{fig:valley-occ-v2-c1}
\end{figure}

\begin{figure}[ht]
\includegraphics[width=\columnwidth]{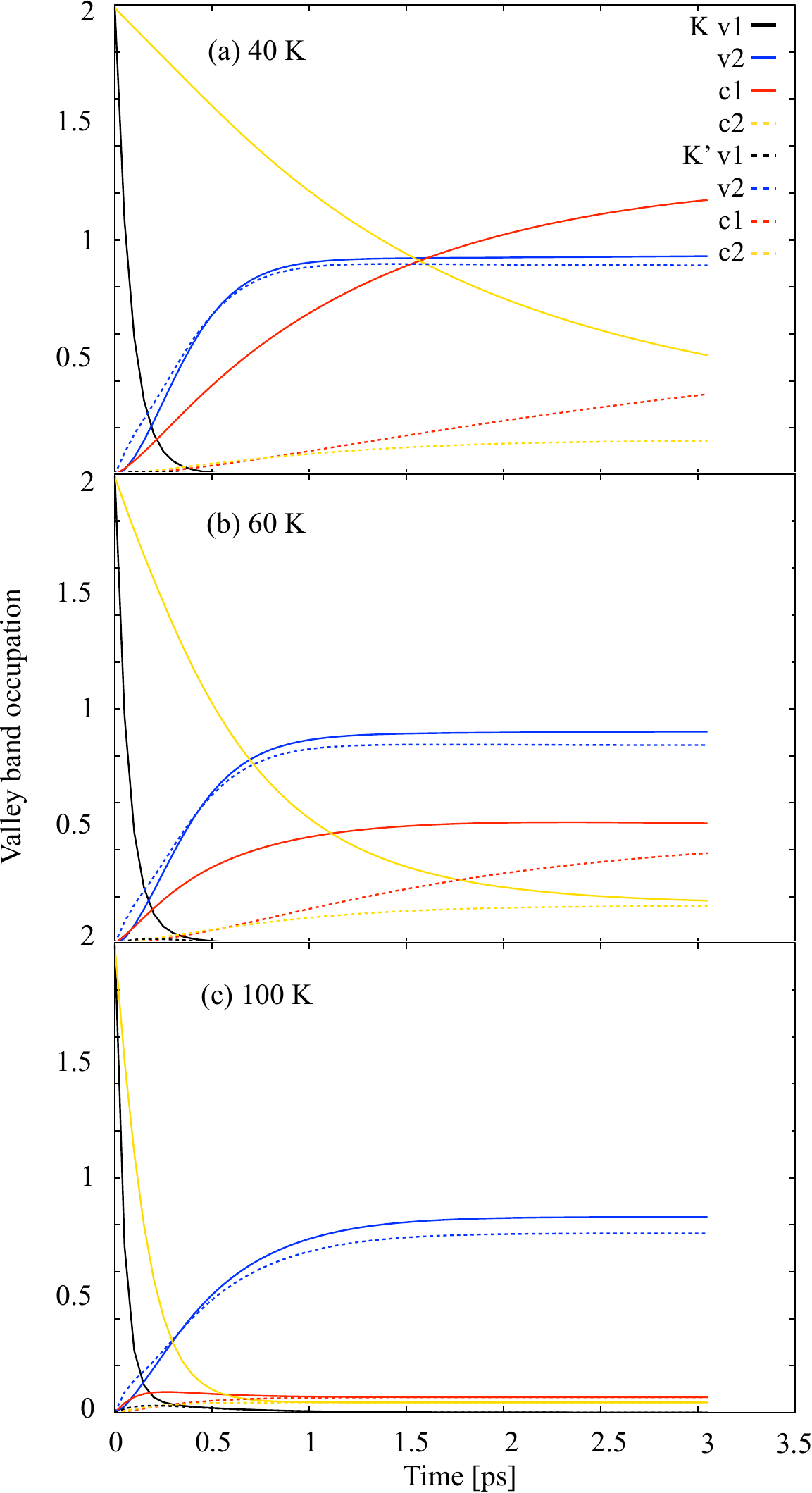}
\caption{Time-evolution of valley electron and hole populations for the artificial excited state in which electrons are moved from v1 to c2 at K, at 40K, 60K, and 100K. Full lines represent the populations in the K valley, while dashed lines show the populations of the corresponding band in the K' valley.}
\label{fig:valley-occ-v1-c2}
\end{figure}

\begin{figure}[ht]
\includegraphics[width=\columnwidth]{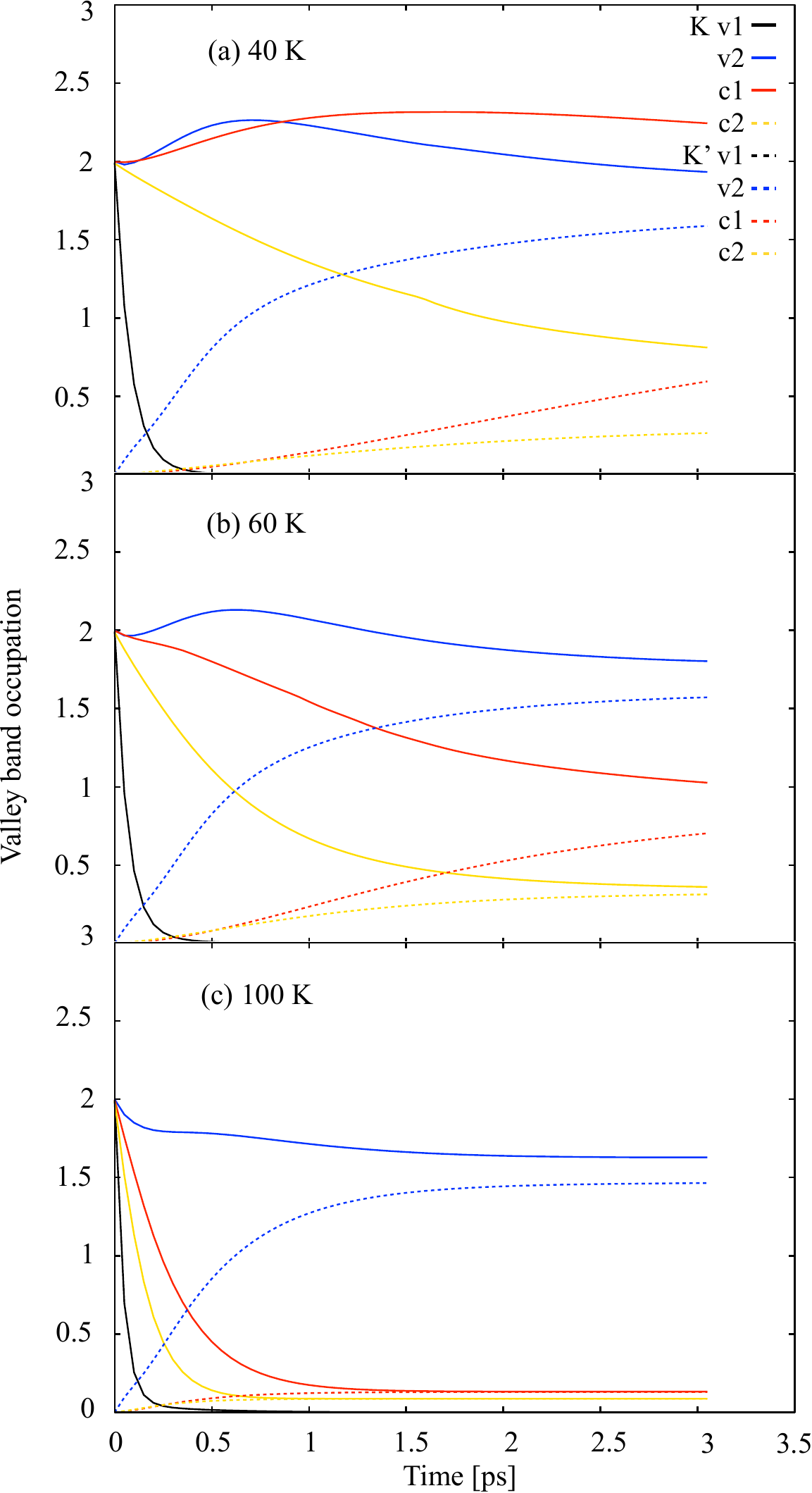}
\caption{Time-evolution of valley electron and hole populations for the artificial excited state in which electrons are moved from v2 to c1, and from v1 to c2 at K, at 40K, 60K, and 100K. Full lines represent the populations in the K valley, while dashed lines show the populations of the corresponding band in the K' valley.}
\label{fig:valley-occ-all}
\end{figure}

\begin{figure}[ht]
\includegraphics[width=\columnwidth]{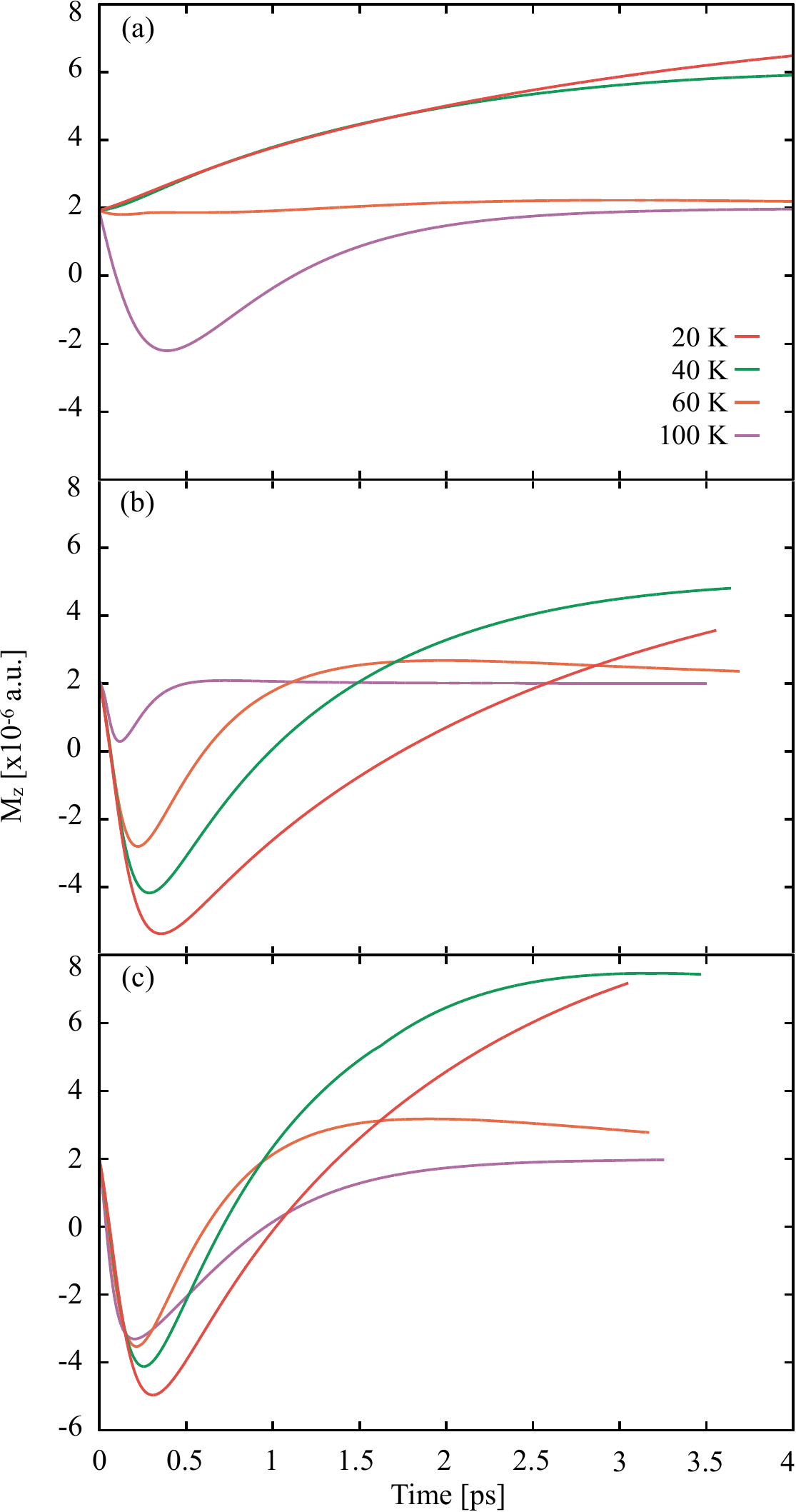}
\caption{(Time evolution of the magnetisation for the three test cases where only the electron-phonon interaction was active. (a) electrons moved from v2 to c1; (b) electrons moved from v1 to c2; (c) both transitions.}
\label{fig:mag-time-elph}
\end{figure}

In Figs.~\ref{fig:valley-occ-v2-c1}  to~\ref{fig:valley-occ-all} we show the evolution in time of the valley band electron and hole populations for the three test cases with the artificially created excited states. These are evaluated by adding the occupations of each state of a given band that lie in a vicinity of either K or K'. The corresponding evolution in the magnetisation is represented in Fig.~\ref{fig:mag-time-elph}.

In all three cases it is possible to conclude the following: while both the scattering of electrons and holes is affected by temperature, it is the electrons that suffer the greatest changes in scattering rates. At 40 K the dominant scattering mechanism for electrons in the conduction states is phonon induced spin flip, due to $\qq = 0$ phonons. In Fig.~\ref{fig:valley-occ-v2-c1} (a) this visible in the changes of the population of the c1 band at K. Its variation in time is less pronounced than that of the holes in band v2 at K. This is understandable, since in order to move to c2, electrons would need to absorb a phonon, but these are scarcely populated at low temperatures. In Fig.~\ref{fig:valley-occ-v1-c2} (a), as we promote electrons to c2 at K', they can scatter more easily to c1, since this would involve emitting a phonon instead of absorbing one. As the temperature increases, electron scattering to other regions of the Brillouin zone becomes possible. Electrons will then disperse throughout the conduction bands faster than the holes, even staying almost completely outside the K and K' valleys. Hence, the very low populations at 3 ps in Figs.~\ref{fig:valley-occ-v2-c1} (c) and~\ref{fig:valley-occ-v1-c2} (c).

The immediate consequence of this change in the scattering rates of electrons relatively to the holes is visible in Fig.~\ref{fig:mag-time-elph} a). At 20 K and 40 K the holes move faster than the electrons towards K, so the remaining contribution to the magnetisation will come from electrons which remain stuck in c1 at K. Their magnetic and spin quantum numbers ($m = -1$ and $m_z = -1$) results in a overall negative signal for the magnetisation. As the temperature increases it is the holes which contribute to the overall magnetisation with their spin and angular momentum ($m = 0$ and $m_z = -1$), thus leading to an overall positive signal.

Regarding the hole scattering in Fig.~\ref{fig:valley-occ-v1-c2}, it is visible that the holes created in v1 at K scatter almost equally to v2 at both K' and K and that this scattering mechanism is not altered significantly due to changes in temperature\footnote{The differences in the seemingly asymptotic values for hole population in both valleys are a result from hole states being populated outside the integration area used to compute the valley band populations.}. The overall changes in the magnetisation are visible in Fig.~\ref{fig:mag-time-elph} b). At low temperatures electron relaxation is dominated by phonon induced spin flip, with electrons remaining trapped in K. This induces a change from states that are dominantly spin-up to states that are dominantly spin-down, inverting the signal of the magnetisation. At higher temperatures electrons are no longer trapped in K, so the holes are again the only contributors to the magnetisation, which reaches zero as the populations in both valleys equalise. 

The convolution of these two effects, phonon-induced spin-flip and intra-valley scattering, is represented in Fig.~\ref{fig:valley-occ-all} and in Fig.~\ref{fig:mag-time-elph} (c). As stated before, here two electron-hole pairs are created, one in bands v2 and c1, and another in bands v1 and c2, both at K. At 40 K [Fig.~\ref{fig:valley-occ-all} (a)] the intra-valley scattering is the dominant process for electron relaxation, even increasing the population of band c1 at K. The same effect is visible in the hole populations, where the holes created in v1 at K scatter into band v2 at both K and K', thus delaying the point at which both valleys would achieve equal populations of holes. By increasing the temperature, inter-valley scattering becomes more prominent for the holes and they are allowed to balance the populations in both valleys. 

The analysis of the effects from the laser pump and the phonon induced scattering allows us to understand  and summarise the results coming from the full simulation. The laser field will create electron-hole pairs, mainly by promoting electrons from band v2 to band c1. However, due to changes in the screening, the exciton binding energies undergo some renormalisation, thus transferring some electrons from v1 to c2. Carriers will then begin relaxing almost immediately after excitation, due to their interaction with phonons. At low temperatures holes relax faster than electrons, which remain trapped in band c1 at the K valley, either because their were placed there by the laser, or because their underwent spin-flip after being promoted to band c2\footnote{Note that the spin-flip transition is allowed because carriers transition first to states around K', which are not pure spin states. This process is very fast. See the file dynamics-V2toC1.mov for a side-by-side comparison of the carrier dynamics for the V2 to C1 transition at 40 K and 60 K.}.  This then results in electrons being the main contributors to the magnetisation at lower temperatures.
By increasing the temperature, electrons can scatter faster than holes, spreading through the whole Brillouin zone. The magnetisation's signal inverts, and the sample will remain magnetised for as long there is an imbalance between the hole populations in the two valleys. 
Our results compare well with the ones of Ref.~\onlinecite{NanoLetters.17.4549}, where the authors have studied the effects of temperature in the time-dependent Kerr signal for WSe$_2$. In that article, however, there was no study of changes with temperature of the relative speed of scattering of electrons and holes due to phonons, or as to how this would reflect in the changes in the magnetisation's signal.
With respect to our system, the one used in Ref.~\onlinecite{NanoLetters.17.4549} has a different spin composition of the bands at K' (and K). In the \mose case, at K bands are ordered as (down, up, up, down), while in WSe$_2$ the ordering is (down, up, down, up). If we were to label the bands in the same way, with respect to their energies, a circularly polarised laser creates electron-hole pairs in K by promoting electrons from v2 to c2. Taking from our observations, this would mean that at low temperatures the dominant relaxation process for electrons would be phonon-induced spin flip, trapping them in band c1 at K'. 
We have shown that the electron-phonon (and also hole-phonon) interaction can account for the sign changes in the Kerr-signal with temperature and have explained which mechanisms (inter and intra-valley scattering) drive those changes. We are in the same regime as Ref.~\onlinecite{NanoLetters.17.4549}, i.e. low intensity of the laser pump, which means that we can safely ignore the electron-hole exchange mechanism, as this is only dominant at high exciton densities and is mostly temperature independent (the temperature will contribute only via the thermal distribution of excitons).